\documentclass[prl,aps,twocolumn,showpacs,preprintnumbers,floats,amsmath,amssymb]{revtex4}

\usepackage{graphicx}
\usepackage{dcolumn}
\usepackage{bm}

\def \be{\begin{equation}}
\def \ee{\end{equation}}

\begin{document}

\title{Tunable Magnetic Relaxation In Magnetic Nanoparticles}

\author{Xavier Waintal$^1$ and Piet W. Brouwer$^2$}
\affiliation{$^1$CEA, Service de physique de l\'etat condens\'e,
Centre d\'etude de Saclay F-91191 Gif-sur-Yvette cedex, France\\
$^2$Laboratory of Atomic and Solid State Physics,
Cornell University, Ithaca, NY 14853-2501 }

\date{\today}

\begin{abstract} 
We investigate the magnetization dynamics of a 
conducting magnetic nanoparticle weakly coupled to source and drain 
electrodes, under the assumption that all relaxation comes from
exchange of electrons with the electrodes. 
The
magnetization dynamics is characterized by a relaxation time $t_1$, which 
strongly depends on temperature, bias voltage, and gate voltage.
While a direct measure of a nanoparticle magnetization might be difficult,
we find that $t_1$ can be determined through a time resolved transport
measurement.
For a suitable choice of gate voltage and bias voltage, the
magnetization performs a bias-driven Brownian motion regardless 
of the presence of anisotropy.
\end{abstract}
\maketitle

Electrical and magnetic dynamics of itinerant ferromagnets are often
described using a generalization of ``Born-Oppenheimer''
approximation, in which the electrons adjust instantaneously to any
changes in the magnetization. This approximation, which is justified
by the wide disparity of electronic and magnetic time scales, is found
to hold down to the smallest size scales attainable by present-day
nanofabrication techniques. The list of physical phenomena it explains
includes the Giant Magneto Resistance effect~\cite{kn:himpsel1998},
the exchange interaction between different layers in ferromagnetic
multilayers \cite{kn:bruno1995}, and the nonequilibrium spin torque
\cite{kn:slonczewski1996}. 
The first-order effect of a changing
magnetization on the electrons was considered only recently
\cite{kn:tserkovnyak2002}.

Once electronic and magnetic degrees of freedom are separated, a
realistic description of magnetization dynamics requires the inclusion
of a relaxation mechanism. 
A phenomenological description of magnetic relaxation is provided by
the Gilbert damping term in the Landau-Lifschitz-Gilbert equation. 
The
Gilbert damping term represents the magnetic relaxation caused by
spin-orbit scattering, phonons, etc; its microscopic origin is the
subject of ongoing research \cite{kn:cehovin2003}. Additional
magnetic relaxation in thin ferromagnetic films
follows from the emission of spin currents into
normal metals adjacent to the ferromagnet \cite{kn:tserkovnyak2002}.

It is the purpose of this letter to study time-dependent electric and
magnetic properties of a magnetic nanoparticle, weakly coupled to
source and drain reservoirs via tunneling contacts. Such magnetic
nanoparticles have been fabricated and studied recently by Ralph and
coworkers \cite{kn:gueron1999,kn:deshmukh2001}. In the absence of
strong spin-orbit coupling in the nanoparticle or for sufficiently
large conductances of the tunneling barriers, all electronic and
magnetic relaxation occurs
via the exchange of electrons with the leads. This system is
sufficiently simple that magnetic and charge degrees of freedom can be
treated on equal footing and the separation of time scales for charge
and magnetization dynamics can be derived from a microscopic model.
Moreover, because magnetic relaxation takes place through the
exchange of electrons with the leads, the magnetic
dynamics crucially depends on the electric environment (bias voltage
and the voltages on nearby metal gates). This dependence leads to
a tunable magnetization relaxation rate. 
To our knowledge, this is the first
system in which magnetic damping rate can be tuned by a simple gate
voltage. Tunability is an important
asset for potential applications of magnetic nanostructures, since
optimal functioning of a nanomagnetic device requires that damping
rate is matched to other relevant time scales of the system.
Further, as we show below, for certain
values of bias and gate voltages, the electric current creates a
non-equilibrium ``randomization'' process that exceeds relaxation,
causing a random motion of the magnetization vector, even in the
presence of magnetic anisotropy. 

\begin{figure}
\vglue +0.45cm
\includegraphics[width=6cm]{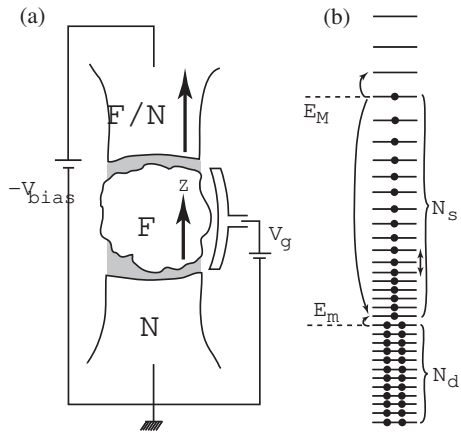}
\caption{\label{fig:system}(a) Schematic of the system under consideration:
a ferromagnetic nanoparticle (F) is connected via tunneling junctions to two 
electrodes, which can be normal (N) or ferromagnetic (F) metals.
A bias voltage $V_{\rm bias}$ is applied across the nanoparticle,
while a gate voltage $V_g$ can be applied on a gate capacitively 
coupled to the particle. (b) 
Structure of the electronic ground state of the nanoparticle. The
ground state has $N_s$ singly occupied levels and Fermi levels 
$E_{\rm M}$ and $E_{\rm m}$ for majority and minority electrons,
respectively.}
\end{figure}

{\it Model.} Starting point of our discussion is a model Hamiltonian
for a ferromagnetic nanoparticle, in which the ferromagnetism arises
from the long-range exchange interaction \cite{kn:kurland2000},
\be
  H_{\rm dot} =\sum_{\mu,\sigma} \epsilon_{\mu} c^{\dagger}_{\mu\sigma} 
  c_{\mu\sigma} - J \vec S \cdot \vec S 
  - \frac{K}{S} S_z^2 + E_C (N-N_{\rm g})^2.
  \label{eq:Hdot}
\ee
Here $c_{\mu\sigma}$ and $c_{\mu\sigma}^{\dagger}$ are annihilation
and creation operators for an electron
with spin $\sigma$ and energy $\varepsilon_{\mu}$, 
\be
  \vec S = \frac{1}{2} \sum_{\mu,\sigma_1,\sigma_2}
  c^{\dagger}_{\mu\sigma_1} \vec\sigma_{\sigma_1\sigma_2}
  c_{\mu\sigma_2}
\ee
is the total spin of the nanoparticle, 
$\vec\sigma$ being the vector of Pauli matrices, $J$ and $K$ set 
the strength of the exchange interaction and anisotropy, 
respectively, $E_C$ is the charging energy, and $N_{\rm g}$ is
proportional to the gate voltage. Without the anisotropy term, this
model is known as the ``universal Hamiltonian'', which has been
shown to describe non-magnetic metal nanoparticles on energy scales below 
the Thouless energy
\cite{kn:kurland2000,kn:aleiner2002}. (Note, however, that,
unlike the ``universal Hamiltonian'', Eq.\ (\ref{eq:Hdot}) should be
considered a model description only, since ferromagnetism
implies a splitting between majority and minority electrons that 
exceeds the Thouless energy.)

Without exchange interaction and magnetic anisotropy, Eq.\
(\ref{eq:Hdot}) is the basis for the ``constant interaction model'' of
the Coulomb blockade \cite{kn:alhassid2000}.  For sufficiently large
exchange interaction, the ground state of the Hamiltonian
(\ref{eq:Hdot}) is ferromagnetic; it has $N_{\rm s} = 2 S$ singly
occupied single-electron levels and different Fermi levels $E_{\rm
M}$ and $E_{\rm m}$ for majority and minority electrons, respectively,
see Fig.\ \ref{fig:system}b, with $E_{\rm M} - E_{\rm m} = 2 S J$. The
excited many-electron states $|\alpha\rangle$ are characterized by 
occupation numbers of the
single-electron levels $n_\mu = 0,1,2$, the total spin $S$, the $z$
component of the total spin, $S_z$, and an additional degeneracy
parameter if $S$ is not maximal. (The $z$ axis is assumed to be the
easy axis of the magnet.) For low-lying excited states, the identity
$N_{\rm s} = 2 S$ is preserved, since many-electron states with
$N_{\rm s} > 2 S$ have an excitation energy of order $E_{\rm M} -
E_{\rm m}$ \cite{kn:canali2000,kn:kleff2001}. The typical energy
scale for low-lying excited states is of the order of the exchange 
energy $J$.
In the nanoparticles studied in Refs.\
\onlinecite{kn:gueron1999,kn:deshmukh2001}, the
anisotropy $K$ is of order $J$ as well
\cite{kn:canali2000}, so that the magnetic excitations $S_z \rightarrow 
S_z\pm 1$ have energy comparable to that of electron-hole 
excitations.

The time-independent low energy properties of this model have already
been extensively analyzed in Refs.\
\onlinecite{kn:canali2000,kn:kleff2001}. Here we study the time
dependence of the total spin $\vec S$ for the case when the
nanoparticle is connected to source and drain reservoirs,
which can
be normal metals or ferromagnets, see Fig.\ \ref{fig:system}a. If
the leads are ferromagnetic, we assume that the magnezation is
collinear with the easy axis of the nanoparticle. For weak
coupling (level broadening much smaller than temperature and level
spacing), the electronic state of the nanoparticle can be characterized by
the probability $P_{\alpha}$ to find the nanoparticle in many-electron
state $|\alpha\rangle$. The probabilities are governed by a rate 
equation,
\begin{widetext}
\begin{eqnarray}
  \label{eq:master}
  \frac{\partial P_\alpha}{\partial t}
  &=& \sum_{\alpha'} \sum_{\mu}
  \sum_{l=L,R} \sum_{\sigma = \uparrow,\downarrow} \left\{
  \Gamma_{\mu l\sigma} 
  \left|\langle \alpha' | c^{\vphantom{\dagger}}_{\mu\sigma} |
  \alpha \rangle \right|^2
\left[(1 - f_l(E_{\alpha} - E_{\alpha'})) P_{\alpha} -
f_l(E_{\alpha} - E_{\alpha'}) P_{\alpha'}\right] \right. \nonumber \\ 
  && \left. \mbox{} +
  \Gamma_{\mu l\sigma} 
 \left|\langle \alpha' | c^{\dagger}_{\mu\sigma} |
  \alpha \rangle \right|^2
\left[f_l(E_{\alpha'} - E_{\alpha}) P_{\alpha}
- (1 - f_l(E_{\alpha'} - E_{\alpha})) P_{\alpha'} 
\right] \right\}.  
\end{eqnarray}
\end{widetext}
Here, the sum over $\alpha'$ extends over all many-electron
states $|\alpha'\rangle$, whereas the sum over $\mu$ extends over
all single-electron states $\mu$. The distribution functions for
the left and right reservoirs are denoted $f_l$, $l=L,R$, 
whereas $\Gamma_{\mu l \sigma}$ set the tunneling rate through contact
$l$ for majority ($\uparrow$) or minority ($\downarrow$) electrons to the
one-body state $\mu$. Equation~(\ref{eq:master}) generalizes the rate
equation used in the theory of Coulomb
blockade~\cite{kn:beenakker1991}. However, one should note that upon
using an evolution equation for {\it probabilities}, information on 
off-diagonal terms of the density matrix is lost. This means 
that, while we keep track of the magnetization 
along the $z$-axis, the position of the magnetization in the $xy$
plane is not monitored. This prevents the use of this approach to the 
case of ferromagnetic leads with non-collinear magnetizations.

{\em Separation of time scales.}
The magnetic structure of the rate equation is encoded in the
matrix elements $\langle \alpha' | c^{\dagger}_{\mu\sigma} |
  \alpha \rangle$. 
Only matrix elements for which the particle number
$N_{\alpha'} = N_{\alpha} \pm 1$ and the spin 
$S_{\alpha'} = S_{\alpha} \pm 1/2$ are nonzero. 
The spin-dependence of the
corresponding transition rate (matrix element squared)
is proportional to the square of a Clebsch-Gordon
coefficient~\cite{kn:messiah1961}. Close to equilibrium ($S_z \approx
S$), these two relevant Clebsch-Gordon coefficients are of order 
$1$ and $1/\sqrt{S}\ll 1$, for up and down electrons respectively tunneling
on a majority state, and vice versa for a minority state.
As we shall see below, it is this hierarchy that leads to the
separation of time scales for the charge and magnetic degrees of
freedom. 

We now consider Eq.\ (\ref{eq:master}) in the limit where both the
temperature and bias voltage are much smaller than $J$, 
so that only one single-electron state $\mu$ is
involved in transport, and for the case that the system is close to
equilibrium, $n = S - S_z \ll S$. For definiteness, we assume that
$\mu$ is a minority state. Denoting the many-electron state with $N$
electrons, spin $S$ and $S_z=S-n$ by ``$(n,-)$'' and the many-electron
state with $N+1$ electrons, spin $S-1/2$ and $S_z=S-1/2-n$ by
``$(n,+)$'', Eq.\ (\ref{eq:master}) reads
\begin{eqnarray}
  \partial_t P_n^+ &=& \sum_{l=L,R} D_{nl}^{+},\ \
  \partial_t P_n^- = \sum_{l=L,R} D_{nl}^{-},
\end{eqnarray}
where
\begin{eqnarray}
  D_{nl}^{+} &=& \Gamma_{\mu l\uparrow} \frac{n+1}{2S+1}
  \left[ f_{ln}^{\uparrow} P_{n+1}^{-} - 
  (1-f_{ln}^{\uparrow}) P_{n}^{+}\right]
  \label{eq:1d+}
  \nonumber \\ && \mbox{}
  + \Gamma_{\mu l\downarrow} \frac{2S-n}{2S+1}
  \left[ f_{ln}^{\downarrow} P_{n}^{-} - 
  (1-f_{ln}^{\downarrow}) P_{n}^{+}\right], \\
  D_{nl}^{-} &=&
  \Gamma_{\mu l\uparrow} \frac{n}{2S+1}
  \left[(1-f_{ln-1}^{\uparrow}) P_{n-1}^+
  - f_{ln}^{\uparrow} P_{n}^{-} \right]
  \nonumber \\ && \mbox{}
  + \Gamma_{\mu l\downarrow} \frac{2S-n}{2S+1}
  \left[(1-f_{ln}^{\downarrow}) P_{n}^+ 
  - f_{ln}^{\downarrow} P_{n}^- \right],  
\label{eq:1d-}
\end{eqnarray}
and we defined $f_{l n \downarrow}
= f_l(E_{n}^{+} - E_{n}^{-})$, and $f_{l n \uparrow}
= f_l(E_{n}^{+} - E_{n+1}^{-})$. The structure of the rate equation
is similar if transport
is facilitated by a majority state, or if several one-electron states
contribute to transport. In all cases, one retains the special
hierarchy of transition rates that is present in Eqs.\
(\ref{eq:1d+}) and (\ref{eq:1d-}): $n=S-S_z$ stays constant in most 
tunneling events (transition rate of order $\Gamma_{\mu l\downarrow}$),
whereas only a small fraction of tunneling events allows $n$
to change by unity (transition rate of order $\Gamma_{\mu l\uparrow} n/(2S+1)$).

To leading order in $n/S$, only the second lines in Eqs.\
(\ref{eq:1d+}) and (\ref{eq:1d-}), corresponding to a process 
where a down spin enters or exits the quantum dot, contributes.
Such a process facilitates charge relaxation, the corresponding
relaxation time scale being
\be 
  t_{\rm c}
  = \left( \Gamma_{\mu L \downarrow} +\Gamma_{\mu R \downarrow}
  \right)^{-1}.
\ee
This is the same time scale as for charge relaxation in a non-magnetic
Coulomb blockaded particle \cite{kn:beenakker1991}. 

The magnetization dynamics appear when we consider terms of 
order $n/S$ in Eqs.~(\ref{eq:1d+}) and (\ref{eq:1d-}). We 
consider the $n$-dependence of the Clebsch-Gordon coefficients in 
Eqs.\ (\ref{eq:1d+}) and (\ref{eq:1d-}), but neglect the (very weak)
$n$-dependence of the Fermi functions $f_{ln}^{\uparrow}$ and
$f_{ln}^{\downarrow}$. Assuming that the charge degree 
of freedom is in equilibrium, the probability for the nanoparticle 
to have spin $S_z = S - n$ is $P_n=P_n^+ + P_n^- = (1+c) P_{n}^{-}$, 
where $c$ is a weighted average of the distribution functions in the 
two leads,
\be
  c =
  \frac{\Gamma_{\mu L \downarrow}
  f_{L}^{\downarrow} + \Gamma_{\mu R \downarrow}
  f_{R}^{\downarrow}}
  {\Gamma_{\mu L \downarrow} + \Gamma_{\mu R \downarrow}}.
\ee
(The ratio $c = P_n^+/P_n^-$ is taken $n$-independent in view of
the comments made above.) The evolution equation for $P_n$ then reads
\begin{eqnarray}
\partial_t P_n &=&
  \gamma_- (n+1) P_{n+1} +\gamma_+ n P_{n-1}
  \nonumber \\ && \mbox{}
  -[\gamma_- n + \gamma_+ (n+1)] P_n,
  \label{eq:Pnrw}
\end{eqnarray}
where we abbreviated
\begin{eqnarray}
  \gamma_{+} &=& \frac{c \Gamma_{\mu L \uparrow} (1 - f_L^{\uparrow})
  + c \Gamma_{\mu R \uparrow} (1 - f_R^{\uparrow})}{2 S + 1},
  \nonumber \\
  \gamma_{-} &=& \frac{(1-c)\Gamma_{\mu L \uparrow} f_L^{\uparrow}
  + (1-c)\Gamma_{\mu R \uparrow} f_R^{\uparrow}}{2 S + 1}.
\end{eqnarray}
In equilibrium
one has $\gamma_- > \gamma_+$, corresponding to the stable 
stationary solution 
\be
  P_n=(\gamma_+/\gamma_-)^n
  (1-\gamma_+/\gamma_-).
\ee 

Equation (\ref{eq:Pnrw}) has a simple interpretation in terms of a 
random walker on a semi-infinite chain that has a probability 
$\gamma_- n$ to step towards the origin and a probability 
$\gamma_+ (n+1)$ to step away from the origin. The difference
$\gamma_- - \gamma_+$ can be identified as the magnetic relaxation
rate $1/t_1$; the sum $\gamma_+ + 
\gamma_-$ corresponds to a ``randomization rate'' for the 
magnetization. The identification $\gamma_- - \gamma_+ = 1/t_1$
becomes manifest when the
time-dependent problem is cast in terms of an evolution equation
for moments of $P_n$ (corresponding to moments of $S_z$).
For the first moment ${\cal M}(t) =\sum_n n P_n(t)$ one finds
\be
\label{eq:llg}
  \partial_t  \mathcal{M}(t) = \gamma_+ - \mathcal{M}(t)/t_1
  ,\ \ t_1^{-1} = \gamma_--\gamma_+.
\ee
which is the equivalent of the Landau-Lifschitz-Gilbert equation for 
our system. 
Note that, while
the charge relaxation rate $1/t_{c}$ depends on 
the tunneling rates only, both $\gamma_-$ and $\gamma_+$ depend strongly 
on temperature, bias voltage, and gate voltage via the Fermi functions 
$f^{\sigma}_{L}$ and $f^{\sigma}_R$.

If a bias voltage is applied on the system, the effect
of the relaxation of $\mathcal{M}(t)$ toward its equilibrium value 
can be inferred from a small change of the measured electric current,
thus allowing for an electric measurement of the magnetic relaxation
time $t_1$,
\begin{eqnarray}
  I &=& \pm e \sum_{n} D_{ln}^{\pm} \nonumber \\
  &=& e \Gamma_{\mu L \downarrow} (f_{L}^{\downarrow} - c) +
  \frac{e \mathcal{M}(t)}{2S+1}\left\{
  \frac{\Gamma_{\mu R \downarrow}\Gamma_{\mu L \uparrow}}
  {\Gamma_{\mu L \downarrow}+ \Gamma_{\mu R\downarrow}}
  (f_{L}^{\uparrow} - c)
  \right.
  \label{eq:currentoft}
  \nonumber \\ && \left. \mbox{}
  - \frac{\Gamma_{\mu L \downarrow}\Gamma_{\mu R \uparrow}}
  {\Gamma_{\mu L \downarrow}+ \Gamma_{\mu R\downarrow}}
  (f_R^{\uparrow} - c) 
 -\Gamma_{\mu L \downarrow} (f_L^{\downarrow} - c)
  \right\},
\end{eqnarray}
[In Eq.\ (\ref{eq:currentoft}) constant terms of order 
$1/(2S+1)$ have been dropped.] 
Whereas additional intrinsic relaxation mechanisms (phonon-magnon 
coupling or spin-orbit coupling), which have not been included in
the above analysis, may alter $t_1$, Eq.\ (\ref{eq:currentoft})
remains valid in all cases.

{\it Non equilibrium induced Browian motion of the magnetization.}
 An interesting limit occurs when
when the bias voltage is bigger than
$E_0^+ - E_0^-$, whereas the temperature is much smaller than
$E_0^+ - E_0^-$. In that case, one may set $f^L_{\downarrow
n}=f^L_{\uparrow n} =  1$ while $f^R_{\downarrow n}=f^R_{\uparrow
n} = 0$. For normal metal leads and a symmetric particle-lead
coupling ($\Gamma_{L\sigma} = \Gamma_{R\sigma}$), 
one then obtains that the stationary
solution is given by the detailed balance solution, 
$P_n^+ = P_n^- = \mbox{constant}$,
independent of the magnetization $n$. [This result can be verified
from direct solution of the detailed balance equation associated with
Eq.\ (\ref{eq:master})]. This total randomization of the 
magnetization occurs irrespectively of the presence of the 
anisotropy: although the
(anisotropy) energy of a state increases with $n$, its probability
$P_n$ does not decrease. This remarkable result is closely linked to the
magnetic structure of the master equation: The bias voltage serves
as a driving force for magnetization randomization, whereas the
magnetization relaxation is suppressed by Coulomb blockade.
We note that through this mechanism, the particle's magnetic energy 
can increases to a value much above the ground state energy. In a 
real sample, this effect is limited by extra sources of relaxation 
that we have not included ({\em e.g.}, spin-orbit scattering).

\begin{figure}[t]
\vglue +0.45cm
\includegraphics[width=0.9\hsize]{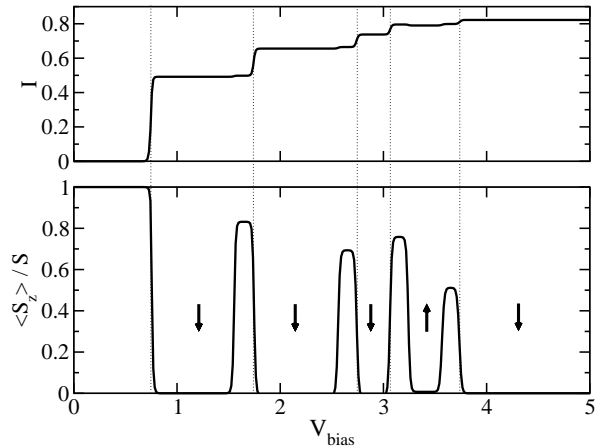}
\caption{\label{fig:I-Sz_Vb} 
Current $I$ (upper) and reduced magnetization $\langle S_z\rangle/S$
(lower) as a function of the bias voltage $V_{\rm bias}$ for a system with
$J=1$, $K=0.1$ and $k_B T=0.01$. The down (up) arrows
correspond to minority (majority) states that become involved in the 
tunneling processes.}
\end{figure}

In Fig.\ \ref{fig:I-Sz_Vb} an example of the bias-voltage
dependence of current and average magnetization is shown, obtained
from numerical solution of the rate equation (\ref{eq:master}).
Except small regions of width $\Delta V_{\rm bias} = K/e$
near current steps, which appear when
a transition $n+1 \rightarrow n$ is allowed while the corresponding 
$n \rightarrow n$ transition is still blocked, one observes
that the magnetization is fully randomized as soon as the bias
voltage exceeds the threshold for current flow. The randomization
effect is suppressed when $K \gtrsim J$. At this point, particle-hole
excitations affect the hierarchy in the structure of the rate
equation (\ref{eq:master}), 
allowing for additional relaxation mechanisms and the breakdown of
detailed balance. (For example, an excited state 
with $N+1$ electrons and $n=0$, which can be accesses from the 
$N$-electron ground state 
with $n=1$, can relax to the $N$-electron ground state 
with $n=0$. However, the opposite process, a direct
transition from the
$N$-electron ground state with $n=0$ to an excited $N+1$-electron 
state with $n=1$, is not allowed.)

{\em Conclusion.} We have studied a simple but 
realistic model of a magnetic dynamics in a magnetic nanoparticle,
for which relaxation occurs via the exchange of electrons with
source and drain reservoirs. For this system,
electric and magnetic properties can be studied on the same footing,
and the separation of electric and magnetic time scales can be
established explicitly. The magnetic dynamics is characterized by
relaxation and randomization rates, which can both be tuned
by bias voltage and gate voltage. A
damping rate well matched to other time scales is crucial for
magnetic dynamics that is both fast and robust. In this sense,
the tunability of the damping rate discovered here may be of use in
future studies of magnetic dynamics on the nanoscale. In addition,
the possibility of measuring $t_1$ through a transport measurement
removes the need of a direct magnetization measurement, which is
difficult for nanoscale magnets.

{\it Acknowledgment. } It is a pleasure to thanks K.\ Mallick,
O.\ Parcollet, D.\ Ralph, and P.\ Roche for 
valuable discussions, as well as P.\ Saldi for the help in drawing
the figures. This work was supported by the NSF under grant 
no.\ DMR 0086509 and by the Packard foundation.


\bibliography{refs,references}


\end{document}